\documentclass[aps,pra,twocolumn,superscriptaddress]{revtex4-1}

\usepackage[utf8]{inputenc}
\usepackage[T1]{fontenc}
\usepackage{amsmath} 
\usepackage{amssymb}
\usepackage{color,graphicx} 
\usepackage{verbatim} 
\usepackage[font=small,labelfont=bf,justification=justified,format=plain]{caption}
\usepackage{subcaption}

\begin{document}
\title{Unruh effect and information flow}

\author{Boris Sokolov}
\email{Electronic address: bosoko@utu.fi}
\affiliation{QTF Centre of Excellence, Turku Centre for Quantum Physics, Department of Physics and Astronomy, University of Turku, FI-20014 Turun Yliopisto, Finland}

\author{Jorma Louko}
\affiliation{School of Mathematical Sciences, University of Nottingham, University Park, Nottingham, NG7 2RD, United Kingdom}
\author{Sabrina Maniscalco}
\affiliation{QTF Centre of Excellence, Turku Centre for Quantum Physics, Department of Physics and Astronomy, University of Turku, FI-20014 Turun Yliopisto, Finland}
\affiliation{QTF Centre of Excellence, Department of Applied Physics, School of Science, Aalto University, FI-00076 Aalto, Finland}
\author{Iiro Vilja}
\affiliation{Turku Centre for Quantum Physics, Department of Physics and Astronomy, University of Turku, FI-20014 Turun Yliopisto, Finland.}

\begin{abstract}
We study memory effects as information backflow for an accelerating two-level detector weakly interacting with a scalar field in the Minkowski vacuum. This is the framework of the well-known Unruh effect: the detector behaves as if it were in a thermal bath with a temperature proportional to its acceleration. Here we show that, if we relax the usual assumption of an eternally uniformly accelerating system, and we instead consider the more realistic case in which a finite-size detector starts accelerating at a certain time, information backflow may appear in the dynamics. Our results demonstrate the existence of a connection between the trajectory of the detector in Minkowski space and the behavior of information flow. This allows us to inspect the Unruh effect under a new light, making use of the latest developments in quantum information theory and open quantum systems. 
\end{abstract}

\maketitle

\section{Introduction}
Quantum field theory predicts that a detector accelerating in empty Minkowski space shall observe a particle bath with a spectrum dependent on the proper acceleration 
of the detector. In particular, if the motion is linear with constant proper acceleration, the particle bath is thermal with a temperature proportional to the acceleration \cite{Unruh1,Crispino}. This extremely minute physical phenomenon is called the Unruh effect. Despite being difficult to detect directly, the effect could prove to be significant in various scenarios such as centripetal acceleration in rotating frames \cite{UnruhRotatingElectrons}. Moreover, there exist several proposals for observing and simulating the Unruh effect in laboratory conditions \cite{SSH,RCPR, VanMat, MFM,PenSud,Cozzella,Jin}.
Since it has not been detected directly, its very existence and meaning have also been questioned \cite{FordOC,MatVan}. From the  theoretical point of view, the Unruh effect is also closely related to Hawking radiation (for a detailed discussion on the subject see Ref. \cite{Crispino}).

Since a constantly accelerated detector experiences an effective thermal background, it is possible to model it as a two-level system interacting with a bosonic environment with a Planckian spectrum. This model has been studied extensively within the framework of open quantum systems theory, both invoking the Born-Markov approximation  \cite{YuZhang,Benatti}, and in more general non-Markovian settings \cite{RavalHuKoks, LinHu, MoustosAnastopoulos}. In all these previous works, both Markovian and non-Markovian, an eternally and constantly accelerating Unruh-DeWitt detector is considered. 

In this paper we focus on the more realistic case of a finite-size detector starting its constant acceleration at a finite time, while still considering weak coupling between the detector and the field. 
The master equation describing the dynamics of the detector in this situation becomes a time-local master equation with time-dependent decay rates which may take temporarily negative values. This time-local structure highlights the departure from the Markovian semigroup dynamics described by the well-known Gorini-Kossakowski-Sudarshan-Lindblad (GKSL) master equation. However, our approach differs from Refs. \cite{RavalHuKoks, LinHu, MoustosAnastopoulos} for two distinct reasons. First, the time-dependent decay rates are always directly dependent on the detector's trajectory, which in our case is different from the standard eternally accelerated case considered in Refs. \cite{RavalHuKoks, LinHu, MoustosAnastopoulos}. Second, we use a modified Wightman function to take into account the detector's profile, as proposed in Ref. \cite{ASatz2} .


During the last decade, a new paradigm in the description of open quantum systems has  emerged. Specifically, a formal and rigorous information-theoretical approach  was introduced and used to define Markovian and non-Markovian dynamics in order to give a clear physical interpretation, as well as an operational definition, to memory effects \cite{BLP,NMDOQS,ReviewRHP,ReviewLHW,ReviewJ}. Markovian dynamics is characterized by a continuous and monotonic loss of information from the open system to the environment while non-Markovian dynamics occurs when part of the information previously lost into the environment comes back due to memory effects, namely information backflow occurs.  

For the system studied in this paper, the time-dependent decay rates appearing in the master equation are obtained from the underlying microscopic Hamiltonian model of system (detector) plus environment (quantum field). Such coefficients are directly linked to the trajectory of the detector in Minkowski space. Interestingly, we have identified the relevant physical parameter ruling the appearance of information backflow and showed under which condition memory effects may occur. This provides new physical insight in the understanding of the Unruh effect and paves the way to the exploration of relativistic quantum phenomena in terms of quantum information exchange between system and environment.

The structure of the paper is as follows. In Sec. II we review the concept of information backflow and how it is related to memory effects and non-Markovian dynamics. In Sec. III. we present our results, namely, (i) we discuss the form of the time-local master equation obtained in the weak coupling limit for a finite-size detector which starts to accelerate at $t=0$; (ii) we study the presence or absence of information backflow and its interpretation, and (iii) we investigate the regions of validity of our approximated master equation by looking at the CP conditions. Finally, in Sec. IV we discuss our results and present conclusions.

\section{Non-Markovianity and information backflow}
The concept of Markovian and non-Markovian stochastic process has a clear and rigorous formulation in the classical domain \cite{breuer-2002}. The extension to quantum processes, however, is not straightforward. Open quantum systems, indeed, may display dynamical features which do not have a classical counterpart, such as recoherence, information trapping, entanglement sudden death and revivals, and so on. For this reason, the generalization of the definition of Markovian/non-Markovian process from classical to quantum is still the subject of an intense debate (for reviews see Ref.\cite{NMDOQS,ReviewRHP,ReviewLHW,ReviewJ}). Generally speaking, there are two approaches to the definition of quantum non-Markovianity. The first one focuses on the properties of the master equation or the corresponding dynamical map, while the second one emphasizes the need of a more physical approach,  identifying memory effects with the occurrence of information backflow. The latter approach does not require the knowledge of the explicit form of either  the master equation or dynamical map, and has been pioneered by Breuer, Laine, and Piilo (BLP), who introduced the now famous BLP non-Markovianity measure \cite{BLP}. In the following we review both perspectives and recall their connection.

\subsection{Non-Markovianity as nondivisibility}
Historically, Markovian open quantum dynamics was identified with the GKSL form of the master equation and was extensively used due to its powerful property of guaranteeing complete positivity (CP), and hence physicality, of the density matrix at all times. A straightforward extension of the GKSL theorem \cite{Lindblad,GKS} to time-local master equations identifies Markovian and non-Markovian dynamics with the properties of the dynamical map $\Phi_\tau: \rho(\tau)=\Phi_\tau \rho(0)$ characterizing the open system evolution. More precisely, the dynamics is said to be Markovian whenever the dynamical map possesses the property of being CP divisible, namely whenever the propagator $V_{\tau,s}$, defined by $\Phi_\tau = V_{\tau,s} \Phi_s$, is CP \cite{Rivas}. This occurs iff the time-dependent decay rates appearing in the master equation are positive at all times $\tau$. On the contrary, non-Markovian dynamics occurs when the dynamical map $\Phi_\tau$ is not CP divisible. This is signaled by the fact that at least one of the time-dependent decay rates of the master equation attains negative values for certain time intervals. 

\subsection{Non-Markovianity as information backflow}
The evolution of a quantum system interacting with its surrounding environment, be it  classical or quantum, relativistic or nonrelativistic, can be described in terms of exchange of energy and/or information between the two interacting parties. While the concept of energy is uniquely defined in quantum systems, a unique definition of information is lacking. Indeed, in principle, there are a number of useful and rigorous choices for quantifying information, and hence information flow, and such choices obviously depend on which "type" of information one is interested in. Quantum information theory deals with the study of information quantifiers, their properties, their dynamics, and their usefulness in quantum computation, communication, metrology and sensing. 

The first attempt to quantify system-environment information flow, and connect it to the Markovian or non-Markovian nature of the dynamics was based on the concept of trace distance between two states $\rho_1$ and $\rho_2$ of an open system,
\begin{equation}
D(\rho_1,\rho_2) = \tfrac{1}{2} \text{tr}\vert \rho_1-\rho_2 \vert .
\end{equation}
The trace distance is invariant under unitary transformations and contractive for CP dynamical maps, i.e., given two initial open-system states $\rho_1(0)$  and $\rho_2(0)$ , the trace distance between the time-evolved states never exceeds its initial value $D[\rho_1(t),\rho_S(t)] \leq D[\rho_1(0),\rho_2(0)]$.

Trace distance is a measure of information content of the open quantum system since it is simply related to the maximum probability $P_D$ to distinguish two quantum states in a single-shot experiment, namely $P_D=\frac{1}{2}[1+D(\rho_1,\rho_2)]$ \cite{NielsenChuang}. Therefore, an increase in trace distance signals an increase in our information about which one of the two possible states the system is in. Following Ref. \cite{BLP}, one can define information flow as the derivative of trace distance as follows:
\begin{equation}
\sigma(t)= \frac{d}{dt} D[\rho_1(t),\rho_2(t)].
\end{equation}

Even though trace distance cannot increase under CP maps, it may not behave always in a monotonic way as a function of time. Specifically, whenever the trace distance decreases monotonically, information flow is negative, meaning that the system continuously loses information due to the presence of the environment. On the other hand, if for certain time intervals information flow becomes positive, then this signals a partial and temporary increase of distinguishability and, correspondingly, a partial recover of information. This information backflow has been proposed as the physical manifestation of memory effects and non-Markovianity. This idea is known as BLP non-Markovianity.

Note that, whenever the dynamical map is BLP non-Markovian, i.e., in presence of information backflow, then it is also CP nondivisible. However, the inverse is not true, namely, there exist systems that are CP nondivisible but BLP Markovian. In general, the concept of nondivisibility and the concept of BLP non-Markovianity, or information backflow quantified by trace distance, do not coincide and their relationship has been the subject of numerous studies (see, e.g., Refs. \cite{ReviewLHW,ReviewJ} for reviews).

\subsection{Connection between nondivisibility and information backflow}

The difference between the concept of CP divisibility and the concept of memory effects due to information backflow, as signaled by an increase of distinguishability, can be overcome if one allows for a more general definition of distinguishability between states. More precisely, the concept of distinguishability based on trace distance is based on the idea of equal probabilities of preparing the two states, i.e., the preparation is uniformly random and there is no prior additional information on which one of the two states is prepared. One can, however, generalize this concept by introducing the Helstrom matrix $\Delta$,
\begin{equation}
  \Delta=p_1 \rho_1 - p_2 \rho_2  
\end{equation}
where $p_1$ and $p_2$ are the prior probabilities of the corresponding states. The information interpretation in terms of the one-shot two-state discrimination problem is valid also in this more general setting \cite{DarekRivas}. 

In more detail, one now considers two states and their corresponding ancilla evolving under the completely positive, trace preserving dynamical map $\Phi_{\tau}$ as follows
\begin{eqnarray}
 \tilde{\rho}_{1,2}(t)= (\Phi_{\tau}\otimes {\cal I}_d) \tilde{\rho}_{1,2}(0),
\end{eqnarray}
with $ \tilde{\rho}_{1,2}$ the combined system-ancilla state, ${\cal I}_d$ the identity map, and $d$ the dimension of the Hilbert space of the system, which in this case is equal to the one of the ancilla.

It has been recently shown in Ref. \cite{DarekRivas} that, for bijective maps, the trace norm of the Helstrom, matrix defined as,
\begin{eqnarray}
 E(t) = |\Delta(t)|=|p_1 \tilde{\rho}_1(t) - p_2\tilde{\rho}_1(t)|
\end{eqnarray}
is monotonically decreasing iff the map is CP divisible. This result has been generalized to nonbijective maps in Ref. \cite{DarekPRL2018}. 
This allows one to interpret lack of CP divisibility in terms of information backflow for system and ancilla, when having prior information on the state of the system, or in our case of the detector.

Finally, one can release the assumption of prior information and prove that, if one uses a $d+1$ dimensional ancilla, then the dynamical map $\Phi_{\tau}$  is CP divisible if and only if the trace distance $D$ decreases or remains constant as a function of time for all pairs of initial system-ancilla states Ref.\cite{Bogna}. Therefore, also in this case, one can interpret the loss of CP divisibility in terms of information backflow for the system-ancilla pair. For further details on the connection between CP divisibility and information backflow we refer the reader to the recent perspective article \cite{ReviewJ}.

In this paper we will specify these approaches to our physical system, and study memory effects and information backflow by looking at the time evolution of the time-dependent decay rates defined by Eq. (\ref{eq:decayrates}). We note that, for the form of master equation considered in this paper, the behavior of the decay rates can be directly connected to the presence or absence of BLP non-Markovianity, and of several other non-Markovianity indicators based on the behavior of other quantifiers of information, as demonstrated by some of the authors of this paper in Ref. \cite{Jose}. Specifically, BLP non-Markovianity can be inferred by the violation of certain sets of inequalities involving the decay rates \cite{Jose}. We will use these results in the follow-up discussions.

\section{Results}
\subsection{The master equation}
In Ref. \cite{Benatti} a microscopic derivation of the master equation describing the dynamics of a two-level detector weakly interacting with a scalar field in the Minkowski vacuum was presented. The derivation relies on the standard Born-Markov approximation \cite{breuer-2002}. An eternally and uniformly accelerated detector parametrized with the proper time, i.e., following the well-known hyperbolic path \cite{Unruh1}, is considered by the authors. Here we relax this unrealistic assumption and consider instead a different trajectory in Minkowski space, assuming that the detector is inertial until a certain time after which it experiences a uniform acceleration. Under these conditions the environment correlation function is not time-translation invariant anymore, and this leads to decay rates which are now time dependent. Moreover, we generalize the description of the detector from pointlike, to finite size. We show in the appendix that with these generalizations, following the same lines of Refs. \cite{Benatti} and \cite{breuer-2002}, the master equation describing the dynamics of the detector takes the form $\dot{\rho} = -i [H_{\mathrm{eff}}, \rho] + \mathcal{L}(\rho)$, where the dissipator $\mathcal{L}$, in the instantaneous rest frame of the detector, is given by
\begin{equation}\label{eq:meLank}
\mathcal{L}(\rho) = \frac{\gamma_1(\tau)}{2} L_1(\rho) + \frac{\gamma_2(\tau)}{2} L_2(\rho) + \frac{\gamma_3(\tau)}{2} L_3(\rho),
\end{equation}
and where the effective Hamiltonian is $H_{\mathrm{eff}} = \omega \sigma_z/2 + \Omega (\tau)$, with $\Omega(\tau)$ a generally time-dependent renormalized frequency.
The dissipator is given by the sum of three terms, $L_i(\rho)$, describing, in order, heating, dissipation and dephasing, and having the following form
\begin{equation}
\begin{split}
L_1(\rho) &= \sigma_+ \rho \sigma_- - \frac{1}{2} \left\{ \sigma_- \sigma_+ ,\rho \right\} \\
L_2(\rho) &= \sigma_- \rho \sigma_+ - \frac{1}{2} \left\{ \sigma_+ \sigma_- ,\rho \right\} \\
L_3(\rho) &= \sigma_z \rho \sigma_z - \rho. \\
\end{split}
\end{equation}
The coefficients $\gamma_1(\tau),\ \gamma_2(\tau)$ and $\gamma_3(\tau)$ are the absorption, emission and dephasing rates, respectively, with the implicit $\omega$ dependence. They are simply related to the proper time ($\tau-$)derivative of the correlation function $F_\tau (\omega )$ through the equations
\begin{equation} \label{eq:decayrates}
\gamma_1(\tau) = 4 \dot{F}_\tau(-\omega),\
\gamma_2(\tau) = 4 \dot{F}_\tau(\omega),\
\gamma_3(\tau) = 2 \dot{F}_\tau(0).
\end{equation}
Note that in this paper we use units $c=\hbar=1$ and Minkowski spacetime signature (+,\,-,\,-,\,-).

For any detector the correlation function is related to the Wightman function 
$W(\tau,\tau')= \langle \phi (\mathtt{x}(\tau ))\phi (\mathtt{x}(\tau'))\rangle$ on the detector worldline  $\mathtt{x}(\tau)$  
as follows \cite{BirrellDavies}:
\begin{equation}
F_{\tau}(\omega) = \int_{\tau_0}^\tau d \tau' \int_{\tau_0}^\tau d \tau''  e^{- i \omega(\tau' - \tau'')} W(\tau',\tau''),
\end{equation}
where $\phi(\mathtt{x})$ is a massless scalar field at Minkowski space point $\mathtt{x} = (t,x,y,z)$. Hence, the proper time derivative $\dot{F}_\tau(\omega)$, for an always-on detector, i.e., for $\tau_0 \rightarrow -\infty$, in its rest frame, reads as
\begin{equation}
\dot{F}_\tau(\omega) = 2 \int_0^{\infty} ds \Re\left( e^{-i\omega s} W(\tau, \tau - s)\right).
\end{equation}

The Wightman function is most easily calculated for a pointlike detector. However, it is not physically realistic and leads to problems e.g. with Lorentz invariance \cite{Schlicht,Satz}.
These problems can be circumvented by assuming that the detector has a finite size instead of being pointlike. The spatial shape of the detector can be defined by the Lorentzian smearing function given in terms of the Fermi coordinates {\boldmath{${\xi}$}} (momentarily normal coordinates) \cite{Schlicht} as
\begin{equation}
f(\text{\boldmath$\xi$})=\frac{1}{\pi^2} \frac{\epsilon^2}{\left( |\text{\boldmath$\xi$}|^2 + \epsilon^2 \right)^2},
\end{equation}
but the detector profile is eventually irrelevant at least if it satisfies some smoothness conditions \cite{ASatz2}. Following the same reference, the transition rate for a pointlike always-on detector is given by
\begin{equation}\label{eq:transrategeneral}
\dot{F}_\tau(\omega) = - \frac{\omega}{4 \pi} + \frac{1}{2 \pi^2}\int_0^\infty \mathrm{d}s \left( \frac{\cos(\omega s)}{(\Delta \mathtt{x})^2} + \frac{1}{s^2}\right),
\end{equation}

while the transition rate for a finite-size detector of characteristic size $\epsilon$ is given by (10) with 
\begin{equation}\label{eq:W2}
W(\tau,\tau') = \frac{-1/4\pi^2}{\left( \mathtt{x}(\tau) - \mathtt{x}(\tau') - i \epsilon \left( \dot {\mathtt{x}}(\tau ) - \dot {\mathtt{x}}(\tau') \right)\right)^2},
\end{equation}
where $\Delta \mathtt{x} := \mathtt{x}(\tau)-\mathtt{x}(\tau-s)$.

This finite-size correlator is more physical, it appears to have much more regular properties, and is therefore used in our study. 

In this paper we consider a detector at rest for $\tau \leq 0$ and uniformly accelerated for $\tau > 0$, following the path given by 

\begin{equation}
\begin{split}
t(\tau) &= \theta(-\tau) \tau + \theta (\tau) \alpha \sinh \left( \frac{\tau}{\alpha} \right),\\
x(\tau) &= \alpha \theta(-\tau) + \alpha \theta(\tau) \cosh \left( \frac{\tau}{\alpha} \right),\\
y(\tau) &= z(\tau)=0,
\end{split}
\end{equation}
where the proper acceleration experienced by the detector is $1/\alpha$, and   $\theta(\tau)$ is the Heaviside step function. 

These more realistic assumptions allow us to perform calculations and obtain explicit expressions for the decay rates. By inserting Eq. (\ref{eq:W2}) and the path into Eq. (\ref{eq:transrategeneral}) we obtain
\begin{equation} \label{eq:impo1}
\begin{split}
2 \pi \alpha \dot{F}_{\bar{\tau}}(\bar{\omega}) =& \frac{\bar{\omega}}{e^{2 \pi  \bar{\omega}} - 1} + \Delta \dot{F}_{\bar{\tau}}(\bar{\omega}) \\
\equiv& \frac{\bar{\omega}}{e^{2 \pi \bar{\omega}} - 1}\\
& +  \frac{1}{\pi} \int_{\bar{\tau}}^\infty \mathrm{d}\bar{s} \cos(\bar{\omega} \bar{s}) \left( \frac{1}{\left( \Delta \mathtt{x}\right)^2_> } - \frac{1}{\left( \Delta \mathtt{x}\right)^2_< }\right),
\end{split}
\end{equation}
where
\begin{equation} \label{eq:impo2}
\begin{split}
\left(\Delta \mathtt{x}\right)^2_> &:= -\left( \sinh(\bar{\tau}) - (\bar{\tau} - \bar{s})\right)^2 + \left( \cosh (\bar{\tau}) - 1\right)^2 \\
\left(\Delta \mathtt{x}\right)^2_< &:= -4 \sinh^2(\bar{s}/2),
\end{split}
\end{equation}
with $\bar{\omega} = \omega \alpha$,
$\bar{\tau} = \frac{\tau}{\alpha}$ and $\bar{s} = \frac{s}{\alpha}$.

For negative times $\bar\tau<0$  the rate of an inertial detector, $\dot{F}_{\bar{\tau}}(\bar{\omega})= -\frac {\omega}{2\pi}\theta(-\omega)$, is restored reflecting the fact that only emission can happen.
For positive times $\bar\tau>0$ the transition rate is the sum of the Planckian equilibrium part ${\bar{\omega}}/({e^{2 \pi  \bar{\omega}} - 1})$ and a dynamical correction
$\Delta \dot{F}_{\bar{\tau}}(\bar{\omega})$ which tends to zero in the asymptotic limit $\bar\tau \rightarrow\infty$. In this limit we obtain the same Lindblad master equation as in Ref. \cite{Benatti}.

Equations (\ref{eq:impo1}) and (\ref{eq:impo2}) allow us to obtain the expression of the decay rates by means of Eq. (\ref{eq:decayrates}) and thus show their connection with the detector trajectory. We note that the behavior of the decay rates crucially depends on the $\alpha$-multiplied angular frequency $\bar\omega$, and hence on both the detector energy $\hbar \omega$ and the proper acceleration; in particular, for fixed $\omega$, larger values of $\bar \omega$ correspond to smaller proper acceleration, i.e. smaller deviation from the inertial system. Also, since the proper acceleration is proportional to the effective Unruh temperature $T_U$, $\bar{\omega}$ can be seen as the ratio between the detector energy  and the effective bath thermal energy $k_B T_U$. We will see that this parameter rules the appearance of information backflow in the Unruh effect.


\begin{figure}
\begin{center}
\vspace{1cm}
\includegraphics[trim={52 0 0 0},clip,width=0.48\textwidth]{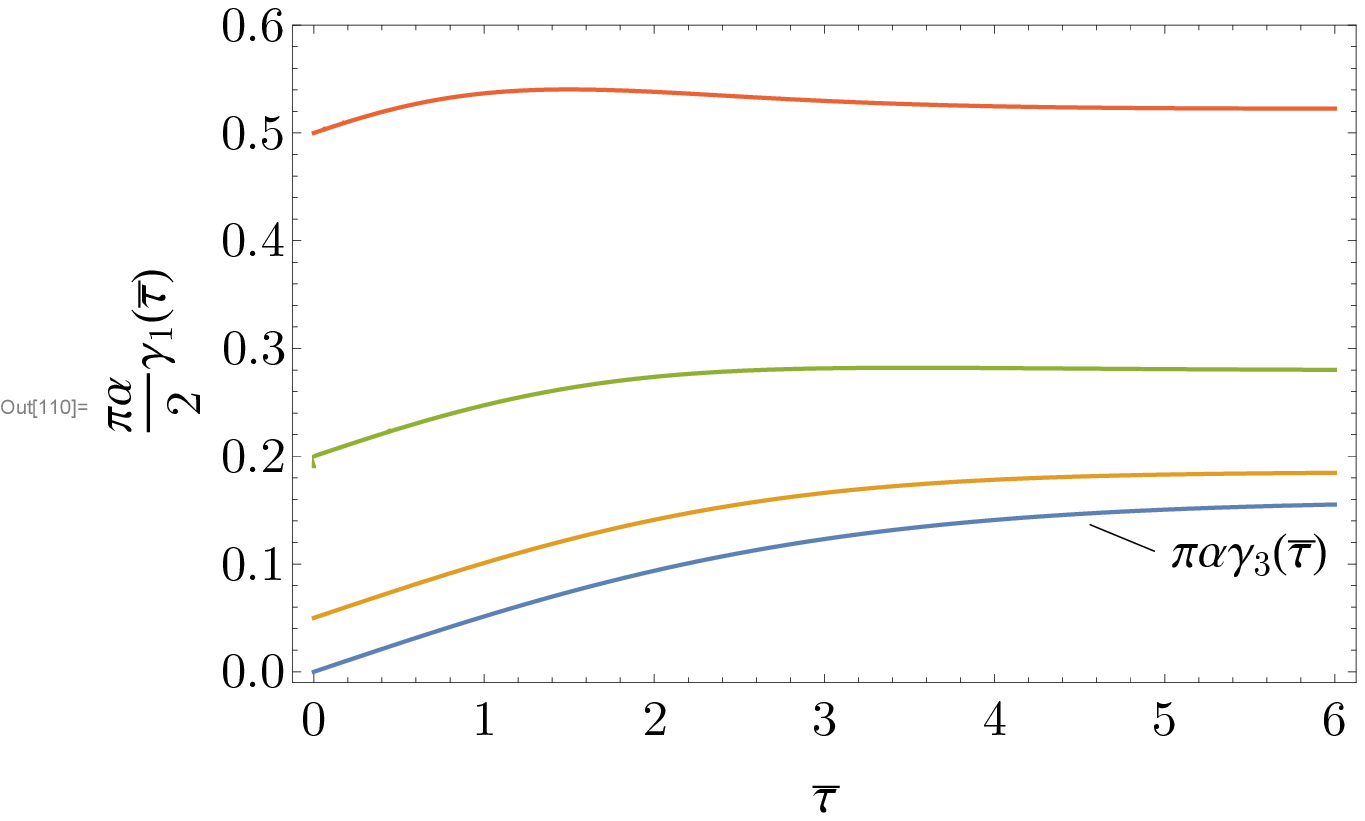}
\caption{Absorption rate $\gamma_1(\bar{\tau})$ for $\bar{\omega} = 0.50\ {\rm (red)}, 0.20\ {\rm (green)}, 0.05\ {\rm (yellow)}$ and dephasing rate $\gamma_3(\bar{\tau})$. \label{fig:wightman8_plots_v118_gam1_050_020_005_gam3}}
\end{center}
\end{figure}

\subsection{Decay rates and information backflow}
In this section we analyze in detail the behavior of the time-dependent decay rates with the aim of understanding the time evolution of information exchange between system and environment. We recall that, if at least one of the coefficients becomes negative at some time, then the map is not CP divisible and therefore  information flows back into the system-ancilla pair.  However, the system can still be BLP Markovian, meaning that there is no information backflow into the system only, but information does return to a larger Hilbert space which includes an ancilla living in a Hilbert space of dimension $d$ (prior information on the state present) or $d+1$ (no prior information on the state present). 

The dephasing rate can be calculated explicitly and has the form
\begin{equation}\label{eq:gamma3}
\pi \alpha \gamma_3(\bar{\tau})  = \frac{1}{2 \pi}\frac{\bar{\tau} - \sinh({\bar{\tau}})}{1 - \cosh({\bar{\tau}})}.
\end{equation}
From this equation we see that $\gamma_3(\bar{\tau})$ is always non-negative for our system. The absorption and emission rates, defined for $\bar{\omega} \neq 0$, require numerical approaches. 
In Fig (\ref{fig:wightman8_plots_v118_gam1_050_020_005_gam3}) we plot sample curves of the absorption and dephasing rates $\gamma_1(\bar\tau)$ and $\gamma_3(\bar\tau)$ weighed by the inverse acceleration factor $\alpha$. These illustrate by examples our extensive numerical investigations showing positivity of the aforementioned rates for all times.

\begin{figure}
\vspace{1cm}
\includegraphics[trim={48 0 0 0},clip,width=0.48\textwidth]{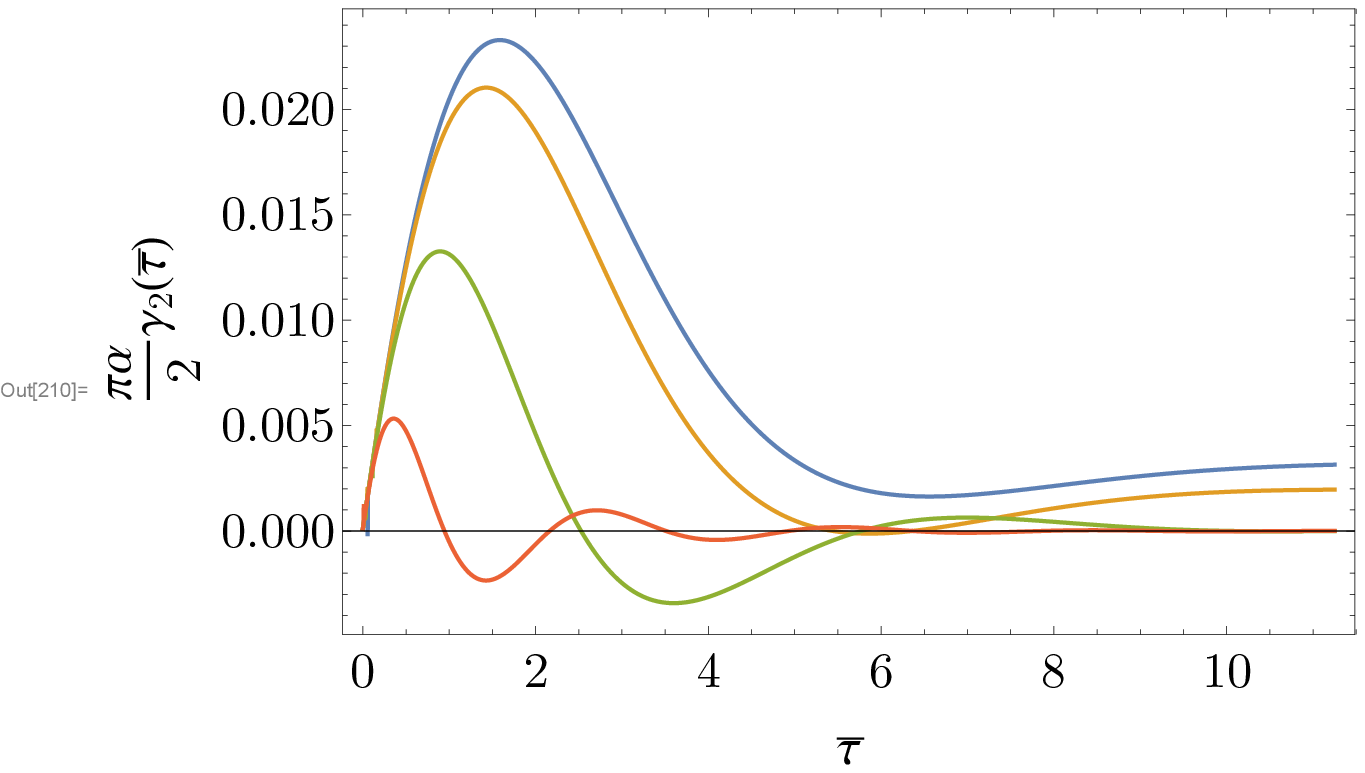}
\caption{Emission rate $\gamma_2(\bar{\tau})$ for $\bar{\omega} = 0.9\ {\rm (blue)}, 1.0\ {\rm (yellow)}, 1.6\ {\rm (green)}, 4.0\ {\rm (red)}$ starting from top, showing non-Markovian regions after $\bar{\omega} \approx 1$  threshold. \label{fig:gamma2}}
\end{figure}

The emission rate $\gamma_2 (\bar{\tau})$ displays a more interesting temporal behavior, since it can attain negative values for $\bar{\omega} \ge 1$, as shown in Fig. (\ref{fig:gamma2}). The parameter $\bar{\omega}$, therefore, controls the transition between CP divisibility and CP nondivisibility, with $\bar{\omega} \approx 1$ the transition value.   In the intervals of time where  $\gamma_2 (\bar{\tau})$ is negative the system-ancilla pair experiences information backflow and memory effects. This happens approximately when the detector energy becomes greater than the thermal energy of the effective bath, i.e., for small Unruh temperatures (or small proper accelerations).

We now conclude our analysis by looking at behavior of other non-Markovianity indicators. In Ref. \cite{Jose} we have established conditions for detecting memory effects using a number of indicators common in the literature, including the BLP non-Markovianity, by means of inequalities involving the decay rates. Since the numerical values of the emission rate are at all times  much higher than those of the absorption rate, as seen from Eq. (\ref{eq:impo1}), the inequalities derived in Ref. \cite{Jose} allow us to conclude immediately that the BLP measure \cite{BLP}, the geometric measure \cite{geometric} and the relative entropy of coherence measure \cite{coherence} do not detect information backflow for any value of $\bar{\omega}$. 

This is consistent with the fact that these three quantities are only indicators of CP nondivisibility; therefore they may not always detect violation of such property. In other words,  in the framework of the system studied, information never returns to the detector only but it will return to a larger system formed by the detector, which interacts with the environment, and an ancilla which does not interact directly with the environment. The ancilla could physically represent, for example, other electronic levels of an atom, if the detector is actually a single atom, or more in general other degrees of freedom which are not explicitly taken into account in the two-state description of the detector and which are not explicitly coupled to the environment.

\subsection{Complete positivity}
We now explore the conditions for complete positivity of the time-local master equation for the Unruh effect discussed in this paper. This is particularly relevant since we know that when the decay rates become negative, and hence the dynamics non-Markovian, we cannot rely anymore on the GKSL theorem to guarantee physicality (i.e., complete positivity) of the solution of the master equation. 
 
In  Ref. \cite{Lankinen} necessary and sufficient conditions for complete positivity for a master equation such as the one here considered has been derived. These conditions are expressed in terms of four inequalities involving the decay rates. By using these inequalities it is straightforward to see that, since in our case $\gamma_3(\tau) > 0$ (Eq. (\ref{eq:gamma3})) at all times, the condition $\tilde{\Gamma}(\tau) = \int_0^\tau ds \gamma_3(s) \geq 0$ is always satisfied. Therefore in our system the complete positivity conditions reduce to the simpler positivity conditions, given by
\begin{equation}\label{eq:cp_conditions}
\begin{split}
P_1(\tau) \equiv e^{-\Gamma(\tau)}\left[ G(\tau)+1\right] \in &\left[0,1\right] \\
P_0(\tau) \equiv e^{-\Gamma(\tau)} G(\tau) \in &\left[0,1\right],
\end{split}
\end{equation} 
where  
\begin{equation}
\begin{split}
\Gamma(\tau) &= \frac{1}{2} \int_0^\tau \mathrm{d}s \left( \gamma_1(s) + \gamma_2(s)\right) \\
G(\tau) &= \frac{1}{2} \int_0^\tau \mathrm{d}s e^{\Gamma(s)}\gamma_2(s).
\end{split} 
\end{equation}

Moreover, $P_{0,1}(\tau)$ can be identified as the ground state probability with initial conditions $P(0)$ equal to 0 or 1, respectively. The positivity conditions of Eq. (\ref{eq:cp_conditions}) can be seen as upper and lower bounds to the ground state probability, respectively. 

Taking the derivative of Eqs. (\ref{eq:cp_conditions}) with respect to $\tau$ we arrive to the same differential equation, with two different boundary values:

\begin{equation}\label{eq:cp_derivatives}
\begin{split}
P'_{1,0}(\tau) &= -P_{1,0}(\tau) \Gamma'(\tau)+\frac{1}{2} \gamma_2(\tau) \\
P_1(0) &= 1 \\
P_0(0) &= 0.
\end{split}
\end{equation}
The upper bounds $P_{0,1}(\tau)\leq 1$ can be studied using Eq. (\ref{eq:cp_derivatives}):

\begin{equation}
\begin{split}
P'_1(0) &= - \frac{1}{2} \gamma_1(0) < 0 \\
P'_0(0) &= \frac{1}{2} \gamma_2(0) > 0,
\end{split}
\end{equation}
as $\gamma_{1,2}(0) = \dot{F}_P (\mp \omega) > 0$, where $\dot F_P(\omega)$ is the Planckian spectrum.  Thus, $P_1(\tau)$ is equal to 1 and decreasing at $\tau=0$ while $P_0(\tau)$ is equal to 0 and increasing at $\tau=0$. Also, both $P_{0}(\tau)$ and $P_{1}(\tau)$ tend to a single finite asymptotic value $\in (0,1)$ as $\tau \rightarrow \infty$, because both $\Gamma'$ and $\gamma_2$ have constant positive asymptotic time limits. 

Suppose now, that $P_{0}(\tau)$ or $P_{1}(\tau)$ is increasing at some time $\tau_1>0$ where it reaches value 1 and would therefore violate complete positivity upper bound for $\tau > \tau_1$. At $\tau = \tau_1$, Eq. \eqref{eq:cp_derivatives} reduces to
\begin{equation}
\begin{split}
P'_{0,1} (\tau_1) &= - \Gamma'(\tau_1) + \frac{1}{2} \gamma_2(\tau_1) \\
&= -\frac{1}{2} (\gamma_1(\tau_1) + \gamma_2(\tau_1)) + \frac{1}{2} \gamma_2(\tau_1)  \\
&= - \frac{1}{2} \gamma_1(\tau_1).
\end{split}
\end{equation}
However, the numerical evidence [see, e.g. Fig. \ref{fig:wightman8_plots_v118_gam1_050_020_005_gam3}]
indicates that $\gamma_1(\tau) > 0 \ \forall \tau > 0$, i.e. the function $P_{0,1}(\tau)$ decreases at the point $\tau_1$ where its value is 1, which is in conflict with the assumption that the function is increasing.
Therefore neither function $P_{0}(\tau)$ nor $P_{1}(\tau)$ can reach the value 1 for any positive time.  Thus, $\forall \tau\geq 0$ we have $P_{0,1}(\tau)\leq 1$, and the upper bounds of the complete positivity conditions are satisfied. 

\begin{figure}
\vspace{1cm} 
\includegraphics[trim={50 0 0 0},clip,width=0.48\textwidth]{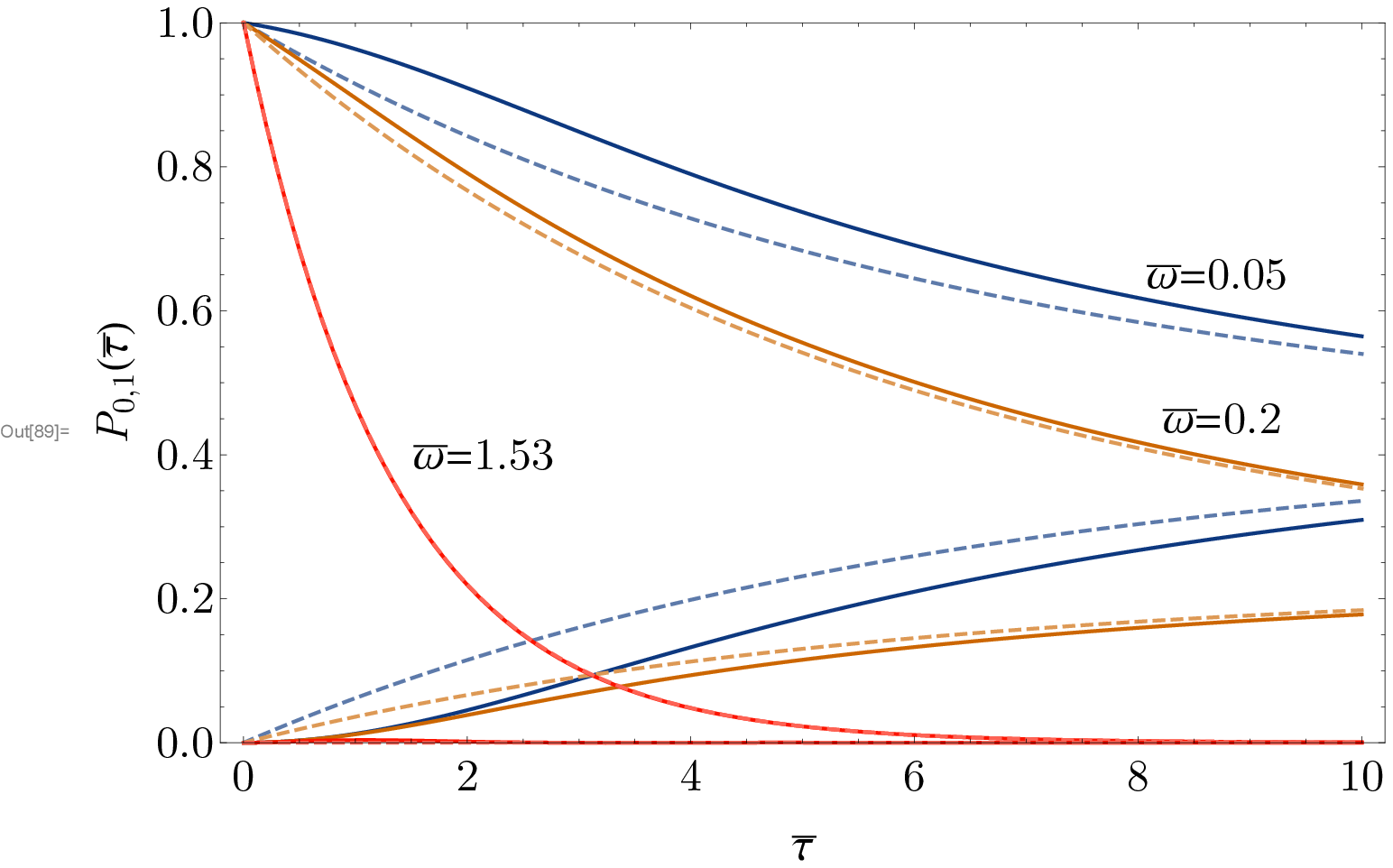}
\caption{$P_0(\bar{\tau})$ and $P_1(\bar{\tau})$. The ground state probabilities for $\bar{\omega}=0.05\ {\rm (blue)}, 0.2\ {\rm (yellow)},\, 1.53\ {\rm (red)}$. Dashed lines represent the Markovian behavior without time-dependent $\Delta \dot F_{\bar\tau}(\bar\omega)$ contribution corresponding to eternally accelerated detector with ground state probability 0 or 1 at $\bar \tau =0$. \label{fig:CP_mark}}
\end{figure}

\begin{figure}
\begin{center}
\vspace{1cm}
\includegraphics[trim={50 0 0 0},clip,width=0.48\textwidth]{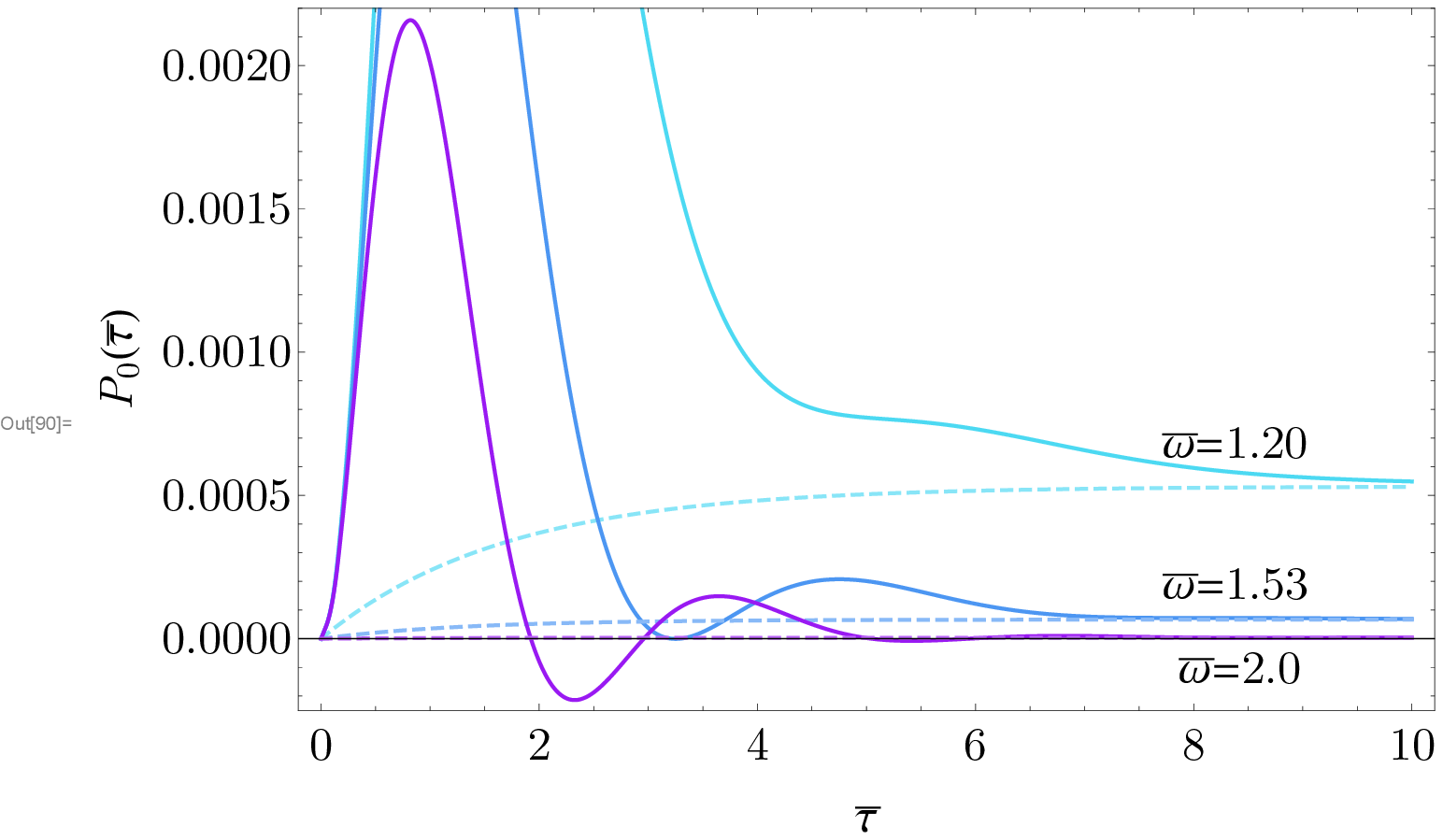}
\caption{The ground state probabilities $P_0(\bar{\tau})$
for $\bar{\omega} = 1.20\ {\rm (light blue)}, 1.53\ {\rm (dark blue)}, 2.0\ {\rm (violet)}$  starting from top, where $P_0(\bar{\tau}) < 0$ indicates CP violation. Dashed lines represent the Markovian behavior without time-dependent contribution.
\label{fig:CP_G}}
\end{center}
\end{figure}

The lower bounds can only be studied numerically. Fortunately, because $P_1 (\tau) > P_0 (\tau)$ only condition $P_0(\tau )\geq 0$ is relevant. In Fig. \ref{fig:CP_mark} we show the dynamics of the ground state probabilities, i.e. functions of the conditions (\ref{eq:cp_conditions}), for some values of 
$\bar{\omega}$. At first sight it seems that the dynamics are completely positive for all times and all considered values of $\bar{\omega}$. However, studying parameter values $\bar{\omega}>1.0$, where the decay rate $\gamma_2 (\tau )$ already exhibits nonpositivity, numerical investigations reveal that the CP condition is violated, i.e.\ $P_0(\bar{\tau})<0$, when $\bar\omega\gtrsim 1.53$ (Fig. \ref{fig:CP_G}). This indicates the breakdown of the approximations used in the derivation of the master equation.

\subsection{Reversed path}
A similar approach to the one in sections above can be directly applied to the case in which the detector decelerates from infinity to rest with a constant negative acceleration rate. This yields the same equation as (\ref{eq:impo1}), now with
\begin{equation}
\begin{split}
(\Delta x)^2_{>} =& -\left( \bar{\tau} - \sinh \left( \bar{\tau} - \bar{s}\right) \right)^2 \\
& + \left(   1 - \cosh \left( \bar{\tau} - \bar{s}\right) \right)^2, \\
(\Delta x)^2_{<} =& \ \bar{s}^2.
\end{split}
\end{equation}

Further analysis, however, shows that the complete positivity conditions fail for all times $\tau > 0$. This is consistent with the fact that the derivation of the master equation with the Lindbladian dissipator in Eq. (\ref{eq:meLank}) assumes complete separability for the initial global state at $\tau = 0$, which is not the case if there has been interaction between the open system and a thermal, $T>0$, environment for all of $\tau < 0$.

\section{Discussion and Conclusions}
When considering the dynamics of the system under study it is worth recalling that, while the accelerated detector undergoes emission and absorption, an inertial detector does not undergo spontaneous excitations. Indeed, more elaborate calculations on the system show that the energy momentum tensor describing the particle content of the space vanishes in any coordinate system, and in particular in the inertial as well as in the rest frame of the accelerated detector \cite{BirrellDavies}. This simply means that the particles detected by the accelerated detector are not real but rather "fictitious" particles. 

The source of energy for the excitation of the accelerating detector is, indeed, its direct coupling to the surrounding vacuum field \cite{Crispino, BirrellDavies, UnWa}. As the detector accelerates, it feels resistance and work is done on it by the external system. The work done not only accelerates the detector but also excites it: to overcome the resistance it is converted into the thermal field affecting the noninertial detector. Thus the energy is not provided by any external particle field but rather originates from the unspecified force keeping the detector in the state of accelerating motion.

In this paper we show that, releasing the eternally accelerated and pointlike detector assumptions, the dynamics may display memory effects and information backflow. The corresponding master equation is time-local with time-dependent decay rates directly linked to the detector worldline. For small enough accelerations the detector keeps memory of the initial time when the acceleration began, and the time evolution becomes non-CP divisible displaying information backflow as defined in Ref. \cite{Bogna}. The same parameter ($\bar{\omega}$) which drives the crossover between the presence or absence of information backflow also controls the range of validity of the master equation, as shown by our study on CP conditions.

Our results shed light on the dynamics of information exchange between the detector and its environment, and specifically on the occurrence of information backflow, in the framework of the Unruh effect. We believe that cross-fertilization between relativistic quantum field theory, open quantum system theory and quantum information theory, may pave the way to a better understanding of a number of open problems by introducing  new tools, diverse approaches and original perspectives. 

\section{Acknowledgements}
The authors acknowledge financial support from the Academy of Finland via the Centre of Excellence program (Project No. 312058) as well as Project No. 287750 and Finnish Academy of Science and Letters. J. L. thanks the University of Turku for hospitality in the early stage of this work. J. L. was supported in part by Science and Technology Facilities Council (Theory Consolidated Grant ST/P000703/1). B. S. thanks the Jenny and Antti Wihuri foundation for financial support.


\section{Appendix}
\subsection{Microscopic derivation of the master equation}

In the microscopic approach to open quantum systems dynamics we start by modeling the total closed system, whose Hilbert space is $\mathcal{H}_S  \otimes  \mathcal{H}_E$, by means of the microscopic Hamiltonian
\begin{eqnarray}
 H=H_{S} \otimes  {\rm I} _E +H_{E}  \otimes  {\rm I} _S +H_{I},
\end{eqnarray}
 where $H_{S}$ and $H_{E}$ are the free Hamiltonians of the system and of the environment, respectively, and $H_{I}$ is the interaction term. The initial state of the total system is assumed to be separable, i.e. no correlations between system and environment are initially present.
 As the total system is closed, we can write its unitary evolution as 
 
 \begin{equation}
 \varrho_{SE} (\tau) = U(\tau) \, \varrho_{S} (0) \otimes \varrho_{E} \, U^{\dag} (\tau), 
 \end{equation}
 with $U(\tau)= \exp [-i H \tau]$. If we now take the partial trace over the environment in the equation above, we have:

\begin{equation}
\begin{split}
\varrho_{S} (\tau) =& \, {\rm Tr}_E \{ U(\tau) \, \varrho_{S} (0) \otimes \varrho_{E} \, U^{\dag} (\tau)\} \\
\equiv & \, \Lambda_t   \varrho_{S} (0) ,
\end{split}
\end{equation}
where $\Lambda_t$ is the dynamical map. In the following we will describe the assumptions that allow us, starting from a microscopic description of system plus environment, to derive a physically meaningful master equation. 
 
Let us consider the dynamics of the overall density operator $\varrho_{SE}$ given by the von Neumann equation which, in units of $\hbar$ and in the interaction picture, reads as follows
\begin{equation}
\frac{d{\varrho}_{SE} (\tau)}{dt}=-i[H_{I}(\tau),\varrho_{SE} (\tau)], \label{eq:1}
\end{equation} 
where we omit for simplicity of notation the subscript $I$ which we should use to indicate the density matrix in the interaction picture.
The integral form of this equation is 
\begin{equation}
\varrho_{SE}(\tau)= \varrho_{SE}(0) - i \int_{0}^{\tau}ds [H_{I}(s),\varrho_{SE}(s)]. 
 \label{eq:2}
 \end{equation}
Inserting Eq. (\ref{eq:2}) into Eq. (\ref{eq:1}) and taking the partial trace over the environmental degrees of freedom we get
\begin{equation}
\frac{d\varrho_{S}}{dt}(\tau)=-\int_{0}^{\tau}ds\textrm{Tr}_{E}\{[H_{I}(\tau),[H_{I}(s),\varrho_{SE}(s)]]\},
 \label{micro1}
 \end{equation}
where we have assumed $\textrm{Tr}_{B}[H_{I}(\tau),\varrho_{SE}(0)]=0$.
 
We assume now that system and environment are weakly coupled (Born approximation). This approximation amounts to assuming that the correlations established between system and environment are negligible at all times (initially zero), i.e.,
$$
 \varrho_{SE}(\tau)\approx\varrho_{S}(\tau)\otimes\varrho_{E}
 $$
Within this approximation we get a closed integro-differential equation for $\varrho_{S}(\tau)$
\begin{equation}
\frac{d\varrho_{S} (\tau)}{dt}=-\int_{0}^{\tau}ds\textrm{Tr}_{E}\{[H_{I}(\tau),[H_{I}(s),\varrho_{S}(s)\otimes\varrho_{E}]]\}
 \label{micro2}
 \end{equation}
Note that, in the equation above, the future evolution of the system, described by $\frac{d\varrho_{s}}{dt}(\tau)$, depends on the past states of the system $\varrho_{S}(s)$ for times $s < \tau$ through the integral.  
A further simplification to this equation is obtained by assuming that we can replace $\varrho_{S}(s)$ appearing inside the integral with its value at time $\tau$, $\varrho_{S}(\tau)$, which is possible if the density matrix does not change strongly in the interval of time $0 \le s \le \tau$. 
This is the case in many physical situations in which this integrand (or rather that part of it describing the environment correlations) quickly decays to zero after a short characteristic correlation time $\tau_E$. This timescale quantifies the memory time of the reservoir. Hence, if the density matrix of the system does not change sensibly in the correlation time $\tau_E$, then we can approximate $\varrho_{S}(s)$ with $\varrho_{S}(\tau)$ in Eq. (\ref{micro2}). The resulting equation is known as the Redfield equation   
 \begin{equation}
\frac{d\varrho_{S} (\tau)}{dt}=-\int_{0}^{\tau}ds\textrm{Tr}_{E}\{[H_{I}(\tau),[H_{I}(s),\varrho_{S}(\tau)\otimes\varrho_{E}]]\}.
 \label{micro4}
 \end{equation}
Equation (\ref{micro4}) is local in time, i.e., the future evolution of the state of the system does not depend on its past state. However, it still retains memory of the initial state $\varrho_{S} (0)$.\\
Until now we have assumed the density matrix does not change much within the correlation time $\tau_E$. The next step will be to neglect such a change altogether by performing a coarse graining in time. This is mathematically achieved by replacing the upper limit of the integral in Eq. (\ref{micro4}) with $\infty$,
\begin{equation}
\frac{d\varrho_{S}}{dt}(t)=-\int_{0}^{\infty}ds\textrm{Tr}_{E}\{[H_{I}(t),[H_{I}(t-s),\varrho_{S}(t)\otimes\varrho_{E}]]\},
 \label{micro44}
 \end{equation}
where we have replaced for the sake of convenience $s$ with $\tau-s$.
The two-step approximation described in Eqs. (\ref{micro4}) and (\ref{micro44}) is known as the Markov approximation. We say that  Eq. (\ref{micro44}) is derived from a microscopic model under the Born-Markov approximation, i.e., for weak coupling and quickly decaying reservoir correlations (memoryless dynamics).
%

Let us decompose the interaction Hamiltonian $H_{I}$ in terms of operators of the system and of the environment:
 $$
 H_{I}=\sum_{\alpha}A_{\alpha}\otimes B_{\alpha}
 $$
with $A_{\alpha} \, (B_{\alpha})$ Hermitian operators of the system (environment). In our case of a two-level system interacting with a scalar field this can be rewritten as
 $$
 H_{I}=\sum_{\alpha}\sigma_{\alpha}\otimes \phi_{\alpha}
 $$
Let us assume that $H_{S}$ has a discrete spectrum and let us indicate with $\epsilon$ the eigenvalues and with $\Pi(\epsilon)$ the corresponding projectors into the corresponding eigenspace. We define the eigenoperators of the system as follows
\begin{equation} 
\sigma_{\alpha}(\omega)=\sum_{\epsilon'-\epsilon=\omega}\Pi(\epsilon)\sigma_{\alpha}\Pi(\epsilon').
\end{equation}
We can rewrite the interaction Hamiltonian in terms of eigenoperators of $H_{S}$, and then pass to the interaction picture exploiting the fact that the system eigenoperators have a simple time dependency in this picture. The environment operators in the interaction picture are simply given by $\phi_{\alpha}(\tau) = e^{i H_E \tau} \phi_{\alpha}  e^{- i H_E \tau}$.

After some algebra, we can rewrite the master equation in the following form
\begin{equation}
\begin{split}
\frac{d\varrho_{s}}{dt}(\tau)=\sum_{\omega,\omega'}\sum_{\alpha,\beta}e^{i(\omega'-\omega)\tau}&\Gamma_{\alpha\beta}(\omega) [\sigma_{\beta}(\omega)\varrho_{S}(\tau)\sigma_{\alpha}^{\dagger}(\omega')  \\ & -
 \sigma_{\alpha}^{\dagger}(\omega')\sigma_{\beta}(\omega)\varrho_{S}(\tau)]+\textrm{h.c.}
\end{split}
\label{micro5}
 \end{equation}
where we introduced
$$
\Gamma_{\alpha\beta}(\omega)\equiv\int_{0}^{\infty}dse^{i\omega s}\langle \phi_{\alpha}^{\dagger}(\tau)\phi_{\beta}(\tau-s)\rangle ,
$$
with the reservoir correlation functions given by
$$
\langle \phi_{\alpha}^{\dagger}(\tau)\phi_{\beta}(\tau-s)\rangle\equiv\textrm{Tr}_{E}\{\phi_{\alpha}^{\dagger}(\tau)\phi_{\beta}(\tau-s)\varrho_{E}\}.
$$
Such correlation functions are homogeneous in time if the reservoir is stationary, i.e.
$$
\langle \phi_{\alpha}^{\dagger}(\tau)\phi_{\beta}(\tau-s)\rangle=\langle \phi_{\alpha}^{\dagger}(s)\phi_{\beta}(0)\rangle,
$$
however this is not true in our case as the field $\phi$ is not invariant under time translations, which is one of the crucial differences from the time-independent case in Ref. \cite{Benatti}.

We now make the last approximation, known as the secular approximation. First we define $\tau_{S}$ as the characteristic intrinsic evolution time of the system. This timescale is generally of the order of $\tau_{S}\approx|\omega'-\omega|^{-1}, \omega'\ne\omega$. We indicate with $\tau_{R}$ the relaxation time of the open system. If $\tau_{S}\gg\tau_{R}$ we can neglect all the exponential terms oscillating at frequency $|\omega'-\omega| \ne0$ as they oscillate very rapidly (averaging out to zero) over the timescale $\tau_R$ over which $\varrho_{S}$ changes appreciably. We then decompose the environment correlation functions into their real and imaginary parts
$$
\Gamma_{\alpha\beta}(\omega)=\frac{1}{2}\gamma_{\alpha\beta}(\omega)+iS_{\alpha\beta}(\omega),
$$
where, for fixed $\omega$, 
$$
\gamma_{\alpha\beta}(\omega)=\Gamma_{\alpha\beta}(\omega)+\Gamma_{\beta\alpha}^{*}(\omega)=\int_{-\infty}^{+\infty}dse^{i\omega s}\langle \phi_{\alpha}^{\dagger}(\tau-s)\phi_{\beta}(\tau)\rangle,
$$
form a positive matrix and 
$$
S_{\alpha\beta}(\omega)=\frac{1}{2i} [\Gamma_{\alpha\beta}(\omega)-\Gamma_{\beta\alpha}^{*}(\omega)],
$$
form a Hermitian matrix.
With these definitions we finally arrive at the interaction picture master equation
\begin{equation}
\frac{d\varrho_{S}}{dt}(\tau)=-i[H_{LS},\varrho_{S}(\tau)]+\mathcal{L}(\varrho_{S}(\tau))
\label{micro6}
\end{equation}
where
$$
H_{LS}=\sum_{\omega}\sum_{\alpha,\beta}S_{\alpha\beta}(\omega)\sigma_{\alpha}^{\dagger}(\omega)\sigma_{\beta}(\omega)
$$
is a Lamb-Shift term which provides a Hamiltonian contribution to the dynamics and
$$
\mathcal{L}(\varrho_{S})=\sum_{\omega}\sum_{\alpha,\beta}\gamma_{\alpha\beta}\left[\sigma_{\beta}(\omega)\varrho_{S}\sigma_{\alpha}^{\dagger}(\omega)-\frac{1}{2}\{\sigma_{\alpha}^{\dagger}(\omega)\sigma_{\beta}(\omega),\varrho_{S}\}\right].
$$ 
This form of the dissipator (generator of the dynamics) $L$ is known as first standard form. Diagonalizing the real positive matrix $\gamma_{\alpha\beta}(\omega)$ we get the GKSL Markovian master equation
$$
\mathcal{L}(\varrho_{S})=\sum_{\omega}\sum_{\alpha}\gamma_{\alpha}(\omega)\left[\bar{\sigma}_{\alpha}(\omega)\varrho_{S}\bar{\sigma}_{\alpha}^{\dagger}(\omega)-\frac{1}{2}\{\bar{\sigma}_{\alpha}^{\dagger}(\omega)\bar{\sigma}_{\alpha}(\omega),\varrho_{S}\}\right],
$$
where $\{\bar{\sigma}_\alpha\}_{\alpha = 0..3} = \{ I, \sigma_+, \sigma_-, \sigma_z \}$.

\end{document}